\documentclass[aps,prd,preprint,eqsecnum,floats,groupedaddress,nofootinbib]{revtex4-2}
\usepackage{epsfig}  
\usepackage{graphicx}
\usepackage{color}
\usepackage[dvipsnames]{xcolor}
\usepackage{float}
\usepackage{amsfonts}
\usepackage{amsmath}
\usepackage{mathrsfs}
\usepackage{slashed}
\usepackage{soul}
\usepackage{braket}
\usepackage{dsfont}
\usepackage[colorlinks,citecolor=blue,urlcolor=blue,bookmarks=false,hypertexnames=true]{hyperref}
\usepackage{orcidlink}

\large

\begin{document} 

\title{Quantum coherence in neutrino spin-flavor oscillations}

\author{Ashutosh Kumar Alok\orcidlink{0000000321557986}}
\thanks{\textcolor{blue}{Deceased}}
\affiliation{Indian Institute of Technology Jodhpur, Jodhpur 342037, India}

\author{Trambak Jyoti Chall\orcidlink{000000025932278X}}
\email{chall.1@iitj.ac.in}
\affiliation{Indian Institute of Technology Jodhpur, Jodhpur 342037, India}

\author{Neetu Raj Singh Chundawat\orcidlink{0000000300926260}}
\email{chundawat.1@iitj.ac.in}
\affiliation{Indian Institute of Technology Jodhpur, Jodhpur 342037, India}

\author{Shireen Gangal\orcidlink{0000000189556205}}
\email{shireen.gangal@gmail.com}
\affiliation{Centre for Excellence in Theoretical and Computational Sciences (CETACS), University of Mumbai,
Santacruz East, Mumbai 400098, India}

\author{Gaetano Lambiase\orcidlink{0000000175742330}}
\email{lambiase@sa.infn.it}
\affiliation{Dipartimento di Fisica, Universitá di Salerno, Via Giovanni Paolo II, 132 I-84084 Fisciano (SA), Italy}\affiliation{INFN, Sezione di Napoli, Gruppo collegato di Salerno, Italy}

\begin{abstract}
Coherence, which represents the superposition of orthogonal states, is a fundamental concept in quantum mechanics and can also be precisely defined within quantum resource theory. Thus exploring quantum coherence in neutrino oscillations can not only help in examining the intrinsic quantum nature but can also explore their potential applications in quantum information technologies. Previous studies on quantum coherence have focused on neutrino flavor oscillations (FO). However, FO imply that neutrinos have mass, and this can lead to the generation of a tiny but finite magnetic dipole moment of neutrinos through quantum loop diagrams at higher orders of perturbative expansion of the interaction. This electromagnetic property of neutrinos can induce spin flavor oscillations (SFO) in the presence of an external magnetic field and hence is expected to enrich the study of coherence. In this work, we investigate quantum coherence in neutrino SFO with three flavor mixing within the interstellar as well as intergalactic magnetic fields, quantified by the $l_1$ norm and the relative entropy of coherence, and express these measures in terms of neutrino SFO probabilities. For FO, coherence measures can sustain higher values (say, within 50\% of the maximum) over distances of several kilometers, which are relevant for terrestrial experiments like reactor and accelerator neutrinos. However, for SFO, we find that the coherence scale can extend to astrophysical distances, spanning from kiloparsecs to gigaparsecs.
\end{abstract}

\maketitle
\newpage

\section{Introduction}

Within the Standard Model (SM) of electroweak interactions, neutrinos were initially considered massless and interacting solely through weak interactions. However, the observation of neutrino oscillation, confirmed by numerous experiments, indicates otherwise \cite{Pontecorvo:1957cp,Pontecorvo:1967fh,Maki:1962mu,Super-Kamiokande:1998uiq,SNO:2002ziz, KamLAND:2002uet,Super-Kamiokande:2004orf,KamLAND:2004mhv,MINOS:2008kxu,MINOS:2011neo,Maltoni:2004ei,Bilenky:1998dt,Bahcall:1998jt,Gonzalez-Garcia:2007dlo}. This experimental confirmation of neutrino oscillation signifies a fundamental departure from the SM framework, indicating that neutrinos indeed possess mass. This discovery stands as the first evidence pointing towards the physics beyond the SM.

The revelation that neutrinos possess mass opens up intriguing possibilities regarding their electromagnetic properties, notably the emergence of phenomena like the generation of neutrino magnetic moment ($\mu_{\nu}$) through quantum loop corrections \cite{Giunti:2014ixa,Valle:2015pba,Giunti:2015gga}. This introduces the prospect of direct interactions between neutrinos and electromagnetic fields. Moreover, these electromagnetic interactions could extend to interactions between neutrinos and charged particles, potentially influencing a diverse array of astrophysical and particle physics phenomena. 
Within the theoretical framework of the Minimal Extended Standard Model (MESM), which introduces right-handed (RH) neutrinos across three generations as the sole additional $\rm SU(2)$ gauge-singlet fields, the predicted value for the neutrino magnetic moment is estimated to be of the order of $10^{-19} \mu_B$ \cite{Lee:1977tib,Fujikawa:1980yx}, where $\mu_B$ represents the Bohr magneton. The value of neutrino magnetic moment can be enhanced in a number of new physics models.

Several experimental efforts have been dedicated to establishing upper bounds on the values of neutrino magnetic moments. Noteworthy among these are the TEXONO experiment \cite{TEXONO:2006xds}, which explores neutrino properties, the GEMMA reactor neutrino experiment \cite{Beda:2012zz}, which focuses on precise measurements of neutrino-nucleus interactions, and the BOREXINO experiment \cite{Borexino:2017fbd}, leveraging solar neutrinos as a primary source for investigation. These experiments have yielded bounds of approximately $10^{-11} \mu_B$. More recently, the XENON experiment \cite{XENON:2022ltv} has provided the latest upper limit, approximately $10^{-12} \mu_B$. Despite the advancements in experimental precision, it's noteworthy that all these upper limits remain considerably higher than the value predicted by the MESM. Therefore, the possibility of large new physics effects remains viable. Hence, the exploration of neutrino electromagnetic interactions has the potential to emerge as a powerful avenue in the pursuit of a comprehensive understanding of beyond SM physics, see for e.g., \cite{Brdar:2020quo,Alok:2022pdn,Nunokawa:1998vh,SinghChundawat:2022mll,Kopp:2022cug,Zhang:2023nxy,Brdar:2023tmi,Koksal:2023qch,Grohs:2023xwa,Sasaki:2023sza,MosqueraCuesta:2009zza,Bulmus:2022gyz,MammenAbraham:2023psg,Lambiase:2004qk,Kurashvili:2020nwb,Miranda:2019wdy,Miranda:2003yh,Kosmas:2015sqa,Miranda:2020kwy,Li:2024gbw,Kosmas:2015vsa,Pastor:1995vn,Jana:2022tsa,Carenza:2022ngg,Baym:2020riw,Giunti:2023yha,Alok:2023sfr,Denizli:2023rqe,Lambiase:2004kf,    Ayala:2024wgb}. Moreover, the ramifications of these interactions extend far beyond the realm of particle physics, exerting significant influence within the vast expanse of astrophysical environments, where neutrinos traverse through magnetic fields present in both free space as well as dense matter.

Due to their weakly interacting nature, neutrinos hold significant promise for long-distance communication applications. This potential may also span vast distances, encompassing terrestrial ranges \cite{Huber:2009kx,MINERvA:2012rrj} as well as extending to extreme scales such as interstellar and intergalactic distances \cite{Learned:2008gr,Pakvasa:2010zz,Anchordoqui:2017eog,Hippke:2017rqi,Learned:2008yd}. Conventional communication systems relying on electromagnetic interactions may not be viable for such expansive distances, as electromagnetic waves are susceptible to damping or deflection in electromagnetic fields. In this context, quantum coherence is anticipated to be pivotal as neutrino beams, undergoing flavor oscillations (FO), represent not just single states but a superposition of states. The theory of neutrino oscillation hinges on the fundamental assumption that different neutrino states stay coherent during their propagation.

Coherence, which identifies the distinctive aspect of quantum mechanics known as the superposition of orthogonal states, is crucial to the framework of quantum mechanics. It represents the defining characteristic of quantum behavior in a singular system and is foundational to various types of quantum correlations in complex systems. Demonstrating quantum coherence in a state is indicative of its authentically nonclassical nature. 

The advancement of quantum information theory has prompted a new evaluation of quantum physical phenomena as resources that could be utilized to accomplish tasks beyond the capabilities of classical physics. This perspective encourages the formulation of a quantitative theory to describe these resources with mathematical precision. 
In this context, coherence stands as a crucial concept that can be precisely defined within the framework of quantum resource theory.
The level of coherence measures a system's potential for quantum-enhanced applications, including quantum computing, quantum communications, quantum cryptography, metrology, nanoscale thermodynamics and quantum algorithms \cite{DiVincenzo:1999,Kelly:2022jfm,Huttner:1995cg,Braunstein:1995jb,Giovannetti:2004cas,Giovannetti:2006amj,Brandao:2013cgt,Skrzypczyk:2016lnb,Gour:2015bhw,Hillery:2016kaa,Shi:2016icg}. Therefore, coherence, a quintessential element of quantum mechanics, emerges as an indispensable building block for quantum technologies.

Drawing on physical intuition, one might be inclined to define coherence merely as functions of the off-diagonal elements of a density matrix. In \cite{Baumgratz:2014yfv}, the concept of quantum coherence was rigorously established by formulating a quantitative theory that treats coherence as a resource akin to the established approaches for entanglement. They introduced several measures of coherence, including the $l_1$ norm of coherence and the relative entropy of coherence, which are grounded in robust metrics. It is important to note that these measures of coherence are universally applicable across all quantum systems, unlike quantum correlation measures that typically require multiple parties\footnote{Study of quantum correlations in the context of neutrino FO have been a topic of great interest in recent times, see for e.g., \cite{Blasone:2007wp,Blasone:2007vw,Alok:2014gya,Banerjee:2015mha,Formaggio_2016,Fu:2017hky,Naikoo:2017fos,Jha:2021itm,Naikoo:2018vug,Naikoo:2019eec,Ming:2020nyc,Blasone:2021mbc,Blasone:2022iwf,Chattopadhyay:2023xwr,Blasone:2023gau,Caban:2007je,Capolupo:2018hrp, Buoninfante:2020iyr}.}. The $l_1$-norm of coherence is intimately connected to quantum multi-slit interference experiments \cite{Bera:2015wbe} and is employed to investigate the advantages of quantum algorithms \cite{Shi:2016icg,Hillery:2016kaa} whereas the relative entropy of coherence is essential for coherence distillation \cite{Winter:2016bkw}, coherence freezing \cite{Bromley:2014gna}, and determining the secret key rate in quantum key distribution \cite{Ma}.

Consequently, investigating quantum coherence in the realm of neutrino oscillations will not only enable the examination of inherent quantumness within these systems but will also pave the way for exploring their potential applications in quantum information technologies. In \cite{Song:2018bma,Dixit:2019swl,Dixit:2018gjc,Joshi:2019etp}, coherence within the framework of neutrino oscillations was explored. However, the situation may diverge intriguingly due to the electromagnetic properties of neutrinos, particularly the neutrino magnetic moment. Subject to external magnetic fields, neutrinos may undergo spin-flavor oscillations (SFO) precipitated by finite magnetic moments resulting from quantum loop corrections. This situation enriches the study of coherence, introducing the dynamic of spin flips in addition to the typical FO, especially relevant for high-energy astrophysical neutrinos traversing extensive distances through interstellar and intergalactic magnetic fields. It is particularly compelling to explore how the embedded quantumness in such oscillating ultra-high-energy astrophysical neutrinos, as quantified by quantum coherence, is affected due to the fact that it also possesses a finite magnetic moment. 

In this work, we focus on investigating the two most general coherence monotones, the $l_1$ norm of coherence and the relative entropy of coherence in neutrino SFO within the interstellar magnetic field of the Milky Way and beyond, in the intergalactic magnetic field. All galaxies, regardless of their age or type, appear to have a static magnetic field with a strength of a few $\mu\rm G$ or less. It follows from the radio synchrotron measurements that within the Milky Way, the magnetic field averaged over 1 kpc (kilo-parsec) radius is about $6\, \mu\rm G$ \cite{Beck:2000dc}. Furthermore, there is substantial evidence for the existence of intergalactic magnetic fields of similar strengths observed within galaxy clusters.  Interestingly, even within the vast emptiness of cosmic voids, there exists a subtle magnetic field, estimated to be about $10^{-15}$ G in strength, spanning across distances of Mpcs (megaparsecs). The current upper and lower limits on intergalactic magnetic fields are of the order of $10^{-9}$ G \cite{Planck:2015zrl} and $10^{-15}$ G \cite{Tavecchio:2010mk,Taylor:2011bn}, respectively.
We work within the framework of SFO with three flavor mixing for Dirac neutrinos $\nu_{\beta}^{L,R}$, where $\beta = e, \mu, \tau$,  for the initial state $\nu_{\mu}^L$.

The plan of the work is as follows.
In Sec. \ref{sec:nmf}, we provide an explicit description of neutrino SFO and present the SFO transition probabilities required for the subsequent section. Following that, in Sec. \ref{sec:qcn}, we first theoretically derive the two proper measures of quantum coherence in terms of the neutrino SFO probabilities in the initial two subsections. In the final subsection, we demonstrate and interpret the nature of these coherence measures for neutrinos originating both within the Milky Way and from extra-galactic sources. We provide our concluding remarks in the final section, Sec. \ref{sec:con}.

\section{Massive neutrino in an external magnetic field}
\label{sec:nmf}
The effective interaction Lagrangian for the generation of the magnetic dipole moment of a massive neutrino written in terms of both the left and right-chiral parts of the neutrino fields is given by,
\begin{equation}
    \mathcal{L}_{\rm eff}=-\frac{1}{2}\sum_{i,j=1}^{3} \mu_{ij} F_{\mu\nu}\bar{\nu}_{L,i}\sigma^{\mu\nu}\nu_{R,j} + \rm h.c.,
    \label{lag}
\end{equation}
where $i$ and $j$ are mass indices, $\sigma^{\mu\nu}=\frac{i}{2}\left[\gamma^{\mu},\gamma^{\nu}\right]$ is the antisymmetric tensor, $F_{\mu\nu}$ represents the electromagnetic field strength tensor,  $\nu_{L,i}$ is the left-handed SM Dirac neutrino field and $\mu_{ij}$ is the element of the $3\times 3$ magnetic moment matrix in the space of three generations of neutrino states.\\
In Eq. \eqref{lag}, we can clearly see that the generation of neutrino magnetic moment induces a chirality flip that occurs on the incoming/outgoing neutrino fields with mass insertion; this compels us to introduce new fields to represent the right-chiral neutrinos (which are not part of the current electroweak picture of the SM), and one of the most viable options is the introduction of RH $\rm SU(2)$ gauge-singlet fields $\nu_{R,j}$ across the three generations of neutrino states, which forms the basis of the theoretical framework of the MESM.\\
Here, we work under the assumption that the Dirac neutrinos do not possess any transition magnetic moments \cite{Popov:2019nkr,Lichkunov:2020lyf}, which is a valid approximation in the MESM, where the magnetic moment of Dirac neutrinos are given by \cite{Giunti:2014ixa},
\begin{equation}
 \mu_{ij}= \frac{3eG_{\mathrm{F}}}{16 \sqrt{ 2}  \pi^2}\left(m_i + m_j\right) \left(\delta_{ij}-\frac{1}{2} \sum_{l=e,\mu,\tau} U_{l i}^\ast U_{l j} \frac{m_l^2}{m_{W}^2}  \right)\,,
 \label{mom}
\end{equation}
where $U$ stands for the PMNS mixing matrix, $G_F$ is the Fermi constant, $l$ is the lepton flavor index and $m_W$ is the mass of the W gauge boson.
Here we can see that the off-diagonal transition moments are highly suppressed with respect to the diagonal ones due to the presence of the factor ${m_l^2}/{m_{W}^2}$ in Eq. (\ref{mom}).
Within the SM, the diagonal ($i=j$) moments for Dirac neutrinos are hence given by,
\begin{equation}
    \mu_{ii}= \frac{3e G_{\mathrm{F}}m_i}{8 \sqrt{ 2}  \pi^2} \simeq 3.2 \times 10^{-19}\left(\frac{m_i}{1\rm eV}\right)\mu_{B}\,,
\end{equation}
whereas the suppressed off-diagonal ($i\neq j$) Dirac moments can be given as,
\begin{eqnarray}
    \mu_{ij}&=& -\frac{3eG_{\mathrm{F}}}{32 \sqrt{ 2}  \pi^2}\left(m_i + m_j\right) \sum_{l=e,\mu,\tau} U_{l i}^\ast U_{l j} \frac{m_l^2}{m_{W}^2}\nonumber\\&\simeq& -3.8\times 10^{-23}\left(\frac{m_i + m_j}{1\rm eV}\right) \sum_{l=e,\mu,\tau} U_{l i}^\ast U_{l j} \frac{m_l^2}{m_{W}^2}\mu_B \,.
\end{eqnarray}

On the other hand, for Majorana neutrinos, there are two additional diagrams with the lepton number flow anti-parallel to the fermion flow, solving which and following a similar procedure as in the Dirac case, we arrive at the expression of the Majorana magnetic moment given by,
\begin{eqnarray}
 \mu_{ij}^M &=& \frac{3eG_{\mathrm{F}}}{16 \sqrt{ 2}  \pi^2} \left(m_i + m_j\right)
\sum_{l=e,\mu,\tau}\left( U_{l j}^* U_{l i}-U_{l j} U_{l i}^\ast \right) \frac{m_l^2}{m_{W}^2}.
 \label{momM}
\end{eqnarray}
We see in Eq. \eqref{momM} that the Majorana magnetic moment matrix is imaginary and antisymmetric, thus having only off-diagonal elements (transition moments) contributing to the magnetic dipole moment. These moments are anticipated to be of the same order of magnitude in terms of the equivalent Dirac transition magnetic dipole moments. It has been shown in \cite{Kurashvili:2017zab} that the transition magnetic moments affect patterns of neutrino oscillations in interstellar magnetic fields for neutrino energies $\gtrsim 100 $ EeV, which is beyond the considered energy range, hence we don't consider transition moments here.

The massive Dirac neutrino spin eigenstate $|\nu_{i}^{s}\rangle$ propagating in the presence of an arbitrarily oriented magnetic field $\boldsymbol{B}$ solves the Dirac equation \cite{Popov:2019nkr},
\begin{equation}
     \left[\gamma_{\mu}p^{\mu} - m_{i} - \mu_{i}\boldsymbol{\left(\Sigma\cdot B\right)} \right] \nu_{i}^{s}(p) = 0,
     \label{evol}
\end{equation}
where $\mu_i$ stands for the diagonal neutrino magnetic moment and $s\, (= \pm 1)$ being the eigenvalues of the spin operator $\hat{S}_{i}$ that commutes with the Hamiltonian ($\hat{H}_{i}$) in the presence of magnetic field and is given by,
\begin{equation}
    \hat{S}_{i} = \frac{m_{i}}{\sqrt{m_{i}^{2}\boldsymbol{B}^{2} + \boldsymbol{p}^{2}\boldsymbol{B}_{\perp}^{2}}} 
    \left[ \boldsymbol{\left(\Sigma\cdot B\right)} - \frac{i}{m_i} \gamma_{0} \gamma_{5} [\boldsymbol{\Sigma \times \boldsymbol{p}}] \boldsymbol{\cdot} \boldsymbol{B} \right]\,.
    \label{spin-op}
\end{equation}
The Hamiltonian ($\hat{H}_{i}$) representing the system here can be obtained simply by hitting Eq. \eqref{evol} with $\gamma_{0}$ as,
\begin{equation}
    \hat{H}_{i} = \gamma_{0} \boldsymbol{\gamma \cdot p} + \mu_{i}\gamma_{0} \boldsymbol{\left(\Sigma\cdot B\right)} + m_{i} \gamma_{0}\,.
    \label{hamiltonian}
\end{equation}
Here the neutrinos are assumed to be propagating in the $z$-direction, thus their momentum is $\boldsymbol{p} = p_{z}$ and the  magnetic field is given by, $\boldsymbol{B} = (B_{\perp}, 0, B_{\parallel})$. The energy spectrum of neutrino can be expressed as \cite{Popov:2019nkr},
\begin{equation}
    E_{i}^{s} = \sqrt{m_{i}^2 + p^2 + \mu_{i}^{2} \boldsymbol{B}^{2} + 2\mu_{i}s\sqrt{m_{i}^{2}\boldsymbol{B}^{2} + \boldsymbol{p}^{2}\boldsymbol{B}_{\perp}^{2}} }\,.
    \label{en-gen}
\end{equation} 

For ultrarelativistic neutrinos, we can approximate their momenta $p \gg m_i$ and $p \gg \mu_\nu B$, thus  reducing Eq. \eqref{en-gen} to,
\begin{equation}
    E_{i}^{s} \approx p + \frac{m_{i}^{2}}{2p} + \frac{\mu_{i}^{2}\boldsymbol{B}^{2}}{2p} + \mu_{i}sB_{\perp}. 
    \label{energy}
\end{equation}
Further, here within the considered environment for neutrino propagation, given the smallness of the neutrino magnetic moment and the upper bounds on the magnetic fields for a given neutrino mass species, we can assume $\mu_i B\ll m_i$, hence neglecting the contribution from the third term in Eq. \eqref{energy}. Thus, the energy for the neutrinos can be approximated as,
\begin{equation}
    E_{i}^{s} \approx p + \frac{m_{i}^{2}}{2p} + \mu_{i}sB_{\perp}. 
    \label{energyf}
\end{equation}
Since we know that the helicity neutrino mass states $|\nu_{i}^{h}\rangle$ ($h$ represents the handedness) are not stationary states in the presence of a magnetic field, we'll expand them over the neutrino spin states $|\nu_{i}^{s}\rangle$ ($s$ represents the spin) as,
\begin{eqnarray}
     |\nu_{i}^{L}(t)\rangle&=&\sum_{s=-1}^{+1} c_i^s |\nu_{i}^{s}(t)\rangle=\sum_{s=-1}^{+1} c_i^s |\nu_{i}^{s}(0)\rangle e^{-iE_i^s t},\\
     |\nu_{i}^{R}(t)\rangle&=&\sum_{s=-1}^{+1} d_i^s |\nu_{i}^{s}(t)\rangle=\sum_{s=-1}^{+1} d_i^s |\nu_{i}^{s}(0)\rangle e^{-iE_i^s t},
\end{eqnarray}

where the spin eigenstates follow the orthonormality condition 
$\langle \nu_j^{s'}|\hat{P}_i^s|\nu_i^s \rangle=\langle \nu_j^{s'}|\nu_i^s \rangle=\delta_{ss'}\delta_{ij}$ and the quadratic combinations of the coefficients $c_i^s$ and $d_i^s$ are given by the matrix elements of the projection operators $\hat{P}_i^s=\hat{P}_i^{\pm}\equiv|\nu_i^{\pm}\rangle\langle\nu_i^{\pm} |=\frac{(1\pm \hat{S}_i)}{2}$ as,
\begin{equation}
    \langle \nu_i^L|\hat{P}_i^{\pm}|\nu_i^L\rangle = |c_i^{\pm}|^2, \quad
     \langle \nu_i^R|\hat{P}_i^{\pm}|\nu_i^R\rangle=|d_i^{\pm}|^2, \quad
     \langle \nu_i^R|\hat{P}_i^{\pm}|\nu_i^L\rangle=d_i^{\pm \ast} c_i^{\pm}\,.
     \label{uno} 
\end{equation}

These equations can be generalised and written together as, 
\begin{equation}
    C_{is}^{h'h} = \langle\nu_{i}^{h'}(0)|\hat{P}_{i}^{s}|\nu_{i}^{h}(0)\rangle\,, \quad h,h'=\{L,R\}
    \label{coeff}
\end{equation}
where $|c_i^{\pm}|^2=C_{is}^{LL}$, $|d_i^{\pm}|^2=C_{is}^{RR}$, and $d_i^{\pm \ast} c_i^{\pm}=C_{is}^{RL}$. 
From this, we get the amplitude of transition from helicity state $h$ to $h'$ as a plane wave expansion,
 \begin{equation}
     \langle\nu_{i}^{h'}\left( 0 \right)|\nu_{i}^{h}\left( t \right)\rangle = \sum_{s} C_{is}^{h'h} e^{-iE_{i}^{s}t}\,.
    \label{amp}
 \end{equation}  

The expression quantifying the likelihood of a transformation from the initial flavor, denoted as  $\alpha$, with a specific  handedness $h$, to the eventual flavor $\beta$, also with a specific handedness, denoted as $h'$ ($h$'s denote left or right-handedness) is given by,
 \begin{equation}
     P_{\alpha \beta}^{h h'}(t) = |\langle \nu_{\beta}^{h'}(0)|\nu_{\alpha}^{h}(t)\rangle|^{2}\,.
     \label{prob}
 \end{equation}

We know that the neutrino flavour eigenstates $|\nu_{\alpha(\beta)}^{h(h')}\rangle$ are expressed in terms of the neutrino mass eigenstates $|\nu_{i(j)}^{h(h')}\rangle$ through the $3\times 3$ leptonic flavor mixing matrix $U_{\alpha(\beta)i(j)}$ called the Pontecorvo–Maki–Nakagawa–Sakata (PMNS) matrix as,
 \begin{equation}
    |\nu_{\alpha(\beta)}^{h(h')}\rangle=\sum_{i(j)}U_{\alpha(\beta)i(j)}|\nu_{i(j)}^{h(h')}\rangle\,.
    \label{mix}
 \end{equation}
 So, expanding the probability expression in Eq. \eqref{prob} using the neutrino mass eigenstates $|\nu_{i}^{h}\rangle$ and the flavor mixing
 relation Eq. \eqref{mix}, we get our probability of SFO as,
\begin{equation}
     P_{\alpha \beta}^{h h'}(t) = \sum_{i=1}^{3} \big|U_{\beta i}^{\ast}U_{\alpha i} \langle \nu_{i}^{h'}(0)|\nu_{i}^{h}(t)\rangle \big|^{2}\,.
     \label{prob-step}
\end{equation}
Using the transition amplitude from Eq. \eqref{amp}, we can deduce,
\begin{equation}
     P_{\alpha \beta}^{h h'}(t) =\sum_{i,j=1}^{3} \sum_{s,s'=-}^+ U^{*}_{\beta i}U_{\alpha i}U_{\beta j}U^{*}_{\alpha j}(C_{is}^{h'h})(C_{js'}^{h'h})^{*} e^{-i(E_i^s - E_j^{s'})t} \,.
     \label{prob-stepf}
\end{equation}

Now, we use this summation degeneration for the mass and spin indices,
\begin{equation}
 \sum_{i>j;s,s'} +\sum_{s>s';i = j}=\sum_{\{i,j,s,s'\}},
 \label{sum}
\end{equation}
using which we can derive the probability as a sum over all possible mass eigenstates in the following explicable form that is applicable for any set of initial ($\alpha,h$) and final ($\beta,h'$) flavor and helicity indices (states),
\begin{eqnarray}
     P_{\alpha \beta}^{h h'}(t) &=& \delta_{\alpha \beta}\delta_{h h'}
     - 4\sum_{\{i,j,s,s'\}}{\rm{Re}}\big([A_{\alpha\beta}^{h h'}]_{i,j,s,s'}\big)\sin^{2}\left(   \frac{E_{i}^{s}-E_{j}^{s'}}{2}\right)t \nonumber\\
     &+&2\sum_{\{i,j,s,s'\}}{\rm{Im}}\big([A_{\alpha\beta}^{h h'}]_{i,j,s,s'}\big)\sin\left( E_{i}^{s}-E_{j}^{s'}\right)t,
     \label{prob-1}
\end{eqnarray}
 where $[A_{\alpha\beta}^{h h'}]_{i,j,s,s'} = U^{*}_{\beta i}U_{\alpha i}U_{\beta j}U^{*}_{\alpha j} (C_{is}^{h'h})(C_{js'}^{h'h})^{*}$.
The last term in Eq. \eqref{prob-1} entails the CP(charge-parity) violation effects as measured by the leptonic Jarlskog invariant defined as \cite{Esteban:2020cvm},
\begin{equation}
J_{CP}\equiv{\rm{Im}}\big(U^{*}_{\beta i}U_{\alpha i}U_{\beta j}U^{*}_{\alpha j}\big)\propto \sin{\delta_{\rm CP}},
\end{equation}
where we assume the CP violating phase $\delta_{\rm CP}$ to be zero here and hence neglect the aforementioned term.
Also, in Eq. \eqref{prob-1}, we interpret the energy difference $\left( E_{i}^{s}-E_{j}^{s'}\right)$ as the phase of SFO. 
Using Eq. \eqref{energyf}, we can derive this energy difference, and under the assumption that all neutrino states possess an equal magnetic moment (i.e. $\mu_i=\mu_j=\mu_\nu$), the frequency $(E_{i}^{s}-E_{j}^{s'})$ can be given as,
\begin{equation}
    E_{i}^{s}-E_{j}^{s'} = \frac{\Delta m_{ij}^{2}}{2p} + \mu_{\nu}(s-s')B_\perp.
    \label{freq}
\end{equation}
The total frequency mentioned in the above equation can be interpreted as a combination of two distinct contributions, the first part as the vacuum oscillation frequency and the second part as the magnetic field-induced frequency. 

Now in the ultrarelativistic limit ($p\gg m_i$), we know the neutrino helicity states $|\nu_i^h (0)\rangle$ reduces to,
\begin{equation}
    |\nu_i^L (0)\rangle=\frac{1}{\sqrt{2}}\begin{pmatrix}
        0\\-1\\0\\1
    \end{pmatrix}, \quad
    |\nu_i^R (0)\rangle=\frac{1}{\sqrt{2}}\begin{pmatrix}
        1\\0\\1\\0
    \end{pmatrix}
    \label{urs}
\end{equation}

Thus, using the aforementioned states from Eq. \eqref{urs} in the definition in Eq. \eqref{coeff} and neglecting terms of the order $\mathcal{O}\left( \frac{m_i^2}{p^2}\right)$ and higher, the plane-wave expansion coefficients $(C_{is}^{h'h})$ can be reduced to $C_{is}^{LL}\simeq \frac{1}{2}$ and $C_{is}^{RL}\simeq-\frac{s}{2}$. These can be used to deduce the required SFO transition probabilities from Eq. \eqref{prob-stepf} for the initial neutrino flavour state $|\nu_e^L\rangle$ (i.e. $\alpha=e$) as \cite{Lichkunov:2022mjf},
\begin{eqnarray}
    P_{e\beta}^{LL}(t)&=&\frac{1}{4}\sum_{i,j=1}^{3} U^{*}_{\beta i}U_{ei}U_{\beta j}U^{*}_{ej}  \sum_{s,s'=-}^+  e^{-i(E_i^s - E_j^{s'})t},
    \label{pp1}\\
    P_{e\beta}^{LR}(t)&=&\frac{1}{4}\sum_{i,j=1}^{3} U^{*}_{\beta i}U_{ei}U_{\beta j}U^{*}_{ej}  \sum_{s,s'=-}^+ ss' e^{-i(E_i^s - E_j^{s'})t},
    \label{pp2}
\end{eqnarray}
where $\beta=e,\mu,\tau$. These equations can be further reduced using the frequency relation Eq. \eqref{freq} as,
\begin{eqnarray}
P_{e\beta}^{LL}(t) = \cos^2{(\mu_\nu B_\perp t)} \Bigg(\sum_{i=1}^{3} |U_{ei}|^2 |U_{\beta i}|^2 +  2 \sum_{i>j=1}^{3} U^{*}_{\beta i}U_{ei}U_{\beta j}U^{*}_{ej} \cos\left({\frac{\Delta m_{ij}^2}{2p}t}\right) \Bigg)\,,
    \label{pp1f}\\
    P_{e\beta}^{LR}(t) =\sin^2{(\mu_\nu B_\perp t)}   \Bigg(\sum_{i=1}^{3} |U_{ei}|^2 |U_{\beta i}|^2 
    + 2 \sum_{i>j=1}^{3} U^{*}_{\beta i}U_{ei}U_{\beta j}U^{*}_{ej} \cos\left({\frac{\Delta m_{ij}^2}{2p}t}\right) \Bigg)\,,
    \label{pp2f}
\end{eqnarray}
where we can use the three-flavor neutrino oscillation parameters $\Delta m_{21}^{2}= 7.42\times 10^{-5}$\,\ eV$^{2}$ and $|\Delta m_{32}|^{2}\simeq|\Delta m_{31}|^{2}= 2.51\times 10^{-3}$\,\ eV$^{2}$ as per the constraints on the neutrino mass-squared splittings imposed by the current global fits, and for the parameter entries in the PMNS matrix, the sines of the mixing angles are  $\sin^2{\theta_{12}}=0.304$, $\sin^2{\theta_{13}}=0.022$ and $\sin^2{\theta_{23}}=0.573$ \cite{Esteban:2020cvm}.
Thus Eq. \eqref{pp1f} gives us the necessary transition probabilities $P_{ee}^{LL}$, $P_{e\mu}^{LL}$ and $P_{e\tau}^{LL}$ whereas Eq. \eqref{pp2f} gives us $P_{ee}^{LR}$, $P_{e\mu}^{LR}$ and $P_{e\tau}^{LR}$.

Considering $|\nu_\mu^L\rangle$ to be our initial neutrino flavour state, we would have the following transition probabilities,
\begin{eqnarray}
P_{\mu\beta}^{LL}(t) = \cos^2{(\mu_\nu B_\perp t)} \Bigg(\sum_{i=1}^{3} |U_{\mu i}|^2 |U_{\beta i}|^2 +   2 \sum_{i>j=1}^{3} U^{*}_{\beta i}U_{\mu i}U_{\beta j}U^{*}_{\mu j} \cos\left({\frac{\Delta m_{ij}^2}{2p}t}\right) \Bigg)\,,
    \label{pp1fm}\\
    P_{\mu\beta}^{LR}(t) =\sin^2{(\mu_\nu B_\perp t)}   \Bigg(\sum_{i=1}^{3} |U_{\mu i}|^2 |U_{\beta i}|^2 + 2 \sum_{i>j=1}^{3} U^{*}_{\beta i}U_{\mu i}U_{\beta j}U^{*}_{\mu j} \cos\left({\frac{\Delta m_{ij}^2}{2p}t}\right) \Bigg)\,.
    \label{pp2fm}
\end{eqnarray}

Here, from these SFO probability expressions in Eqs. \eqref{pp1f}, \eqref{pp2f}, \eqref{pp1fm} and \eqref{pp2fm}, we can show that when a significant number of SFO cycles are completed for a large enough distance, then the $\cos\left({\frac{\Delta m_{ij}^2}{2p}t}\right)$ term averages to $0$, and the $\cos^2{(\mu_\nu B_\perp t)}$ and $\sin^2{(\mu_\nu B_\perp t)}$ terms average to $\frac{1}{2}$ and thus we're left with, 
\begin{equation}
P_{\alpha\beta}^{LL}(t) = \frac{1}{2}\sum_{i=1}^{3} |U_{\alpha i}|^2 |U_{\beta i}|^2 \,, \quad
    P_{\alpha\beta}^{LR}(t) =\frac{1}{2}\sum_{i=1}^{3} |U_{\alpha i}|^2 |U_{\beta i}|^2 \,.
    \label{pp2favg}
\end{equation}
We can clearly deduce from here that if we sum over $\alpha$ and $\beta$ for all the possible flavors in the Eq. \eqref{pp2favg}, they both reduce to $\frac{1}{2}$ as the summation $\sum_{\alpha,\beta}\sum_{i=1}^{3} |U_{\alpha i}|^2 |U_{\beta i}|^2 = 1$, i.e., 
\begin{equation}
    \sum_{\alpha,\beta}P_{\alpha\beta}^{LL}=\sum_{\alpha,\beta}P_{\alpha\beta}^{LR}=\frac{1}{2} \,,
    \label{avg}
\end{equation}
where the flavor indices $\alpha,\beta=e, \mu, \tau$.

Now, if we use the probability expressions in Eqs. \eqref{pp1f}, \eqref{pp2f}, \eqref{pp1fm} and \eqref{pp2fm} for the two-flavour mixing case by setting $\theta_{13}=\theta_{23}=0$, making the flavor mixing matrix a $2\times 2$ unitary matrix (i.e. the $\rm SU(2)$ rotation matrix) and the frequency in Eq. \eqref{freq} boiled down as $E_{i}^{s}-E_{i}^{s'}=2\mu_{\nu}B_\perp$ for $s>s'$, $E_{i}^{s}-E_{j}^{s'}=\frac{\Delta m_{ij}^{2}}{2p}$ for $i>j$ and $s\geq s'$, $E_{i}^{s}-E_{j}^{s'}=\frac{\Delta m_{ij}^{2}}{2p}$ where $i,j=\{1,2\}$ and $s,s'=\{-,+\}$, then we can easily derive the subclass of solutions \cite{Popov:2018seq,Popov:2019nkr} that will be required for calculation of the measures of quantum coherence as,

{\small\begin{eqnarray}
    P_{ee}^{LL}(t)&=&|\langle \nu_{e}^{L}|\nu_{e}^{L}(t)\rangle|^2=\cos^2{(\mu_\nu B_\perp t)} \left( 1-\sin^2{(2\theta_{12})}\sin^2{\left({\frac{\Delta m_{21}^2}{4p}}t\right)} \right)=P_{\mu\mu}^{LL}(t),
    \label{pllee}\\
    P_{ee}^{LR}(t)&=&|\langle \nu_{e}^{R}|\nu_{e}^{L}(t)\rangle|^2=\sin^2{(\mu_\nu B_\perp t)}\left( 1-\sin^2{(2\theta_{12})}\sin^2{\left({\frac{\Delta m_{21}^2}{4p}}t\right)} \right)=P_{\mu\mu}^{LR}(t),
    \label{plree}\\
    P_{e\mu}^{LL}(t)&=&|\langle \nu_{\mu}^{L}|\nu_{e}^{L}(t)\rangle|^2=\cos^2{(\mu_\nu B_\perp t)} \sin^2{(2\theta_{12})}\sin^2{\left({\frac{\Delta m_{21}^2}{4p}}t\right)}=P_{\mu e}^{LL}(t),
    \label{pllem}\\
    P_{e\mu}^{LR}(t)&=&|\langle \nu_{\mu}^{R}|\nu_{e}^{L}(t)\rangle|^2=\sin^2{(\mu_\nu B_\perp t)} \sin^2{(2\theta_{12})}\sin^2{\left({\frac{\Delta m_{21}^2}{4p}}t\right)}=P_{\mu e}^{LR}(t).
    \label{plrem}
\end{eqnarray}}
In our analysis, we restrict our focus to diagonal magnetic moments for Dirac neutrinos. For two-flavor Dirac and Majorana neutrino SFO in the interstellar medium, the impact of transition magnetic moments was considered in \cite{Kurashvili:2017zab} specifically for ultrahigh energy cosmic neutrinos. However, a more detailed investigation would be required for the three-flavor mixing case in Majorana neutrinos, as it would need to include non-diagonal transition moments \cite{Lichkunov:2020lyf}. We are now equipped with all the transition probabilities required for calculating our measures of quantum coherence in the system of Dirac neutrinos undergoing SFO under a pervasive magnetic field.

\section{Quantifying coherence in neutrino SFO}
\label{sec:qcn}
For a fixed reference basis $\{\ket{i}\}_{\rm i=1,2,..n}$ in the $n$-dimensional Hilbert space, the definition of incoherent states is given as,
\begin{equation}
    \sigma = \sum_{i=1}^{n} \sigma_i \ket{i}\bra{i},
    \label{def}
\end{equation}
where $|\sigma_i|^2$ represents the probabilities of occurrence. As is apparent from this Eq. \eqref{def}, the states (represented by density matrices) that are diagonal in the basis $\{\ket{i}\}_{\rm i=1,2,..n}$ are incoherent, and the states that cannot be expressed in this form are termed coherent.
An $\rm U(1)$-covariant operation that preserves trace (TP) and exhibits complete positivity (CP), is classified as an incoherent operation if it can be expressed as a function of a quantum state $\rho$ as,
\begin{equation} \label{Gaerho}
    \Xi_{\rm ICPTP}(\rho) = \sum_{\ell} K_\ell \rho K_\ell^\dagger\,, 
\end{equation} 
where $\ell$ represents individual outcomes and the operators $K_\ell$, known as incoherent Kraus operators (satisfying $\sum_{\ell}K_\ell^\dagger K_\ell=\mathds{1}$), transform any incoherent state into another incoherent state fulfilling $K_\ell \mathcal{I}  K_\ell^\dagger\, \subset \mathcal{I}$ $ \forall$ $\ell$, with $ \mathcal{I}$
representing the set of incoherent quantum states. Any proper measure of coherence represented by a functional of density operators $C(\rho)$ mapping states to non-negative real numbers should fulfil the conditions listed below: 
\begin{itemize}
    \item \textit{Non-negativity}:  $C(\rho) \geq 0$ for any quantum state  $\rho$, and particularly $C(\rho)$ should vanish on the set of incoherent states, i.e., $C(\sigma) =0$ if and only if $\sigma \in \mathcal{I}$, i.e., iff  $\sigma$  is an incoherent state.

    \item \textit{Monotonicity}:  $C(\rho)$ should not increase under the action of incoherent completely positive and trace-preserving operations, i.e., $C(\Xi_{\rm ICPTP}(\rho)) \leq C(\rho)$ $\forall$ $\Xi_{\rm ICPTP}$. The functionals of density operators satisfying this condition are called ``monotones'' under the action of $\Xi_{\rm ICPTP}$.

    \item  \textit{Strong monotonicity}: The above condition does not account for subselection based on measurement outcomes, so when we retain the measurement outcomes, the monotonicity condition becomes stronger, i.e., $C(\rho)\geq \sum_{\ell}p_{\ell}C(\rho_{\ell})$ $\forall$ $\{K_{\ell}\}$. Here, the state corresponding to outcome $\ell$ is given by $\rho_{\ell}=\frac{K_{\ell}\rho K_{\ell}^{\dagger}}{p_{\ell}}$ where the probability of occurrence of such a state is given by $p_{\ell}=\rm{Tr}(K_\ell \rho K_\ell^\dagger)$. This condition is also termed as \textit{Ensemble} monotonicity.

    \item  \textit{Convexity}\footnote{In the extension of the coherence measures for mixed states, the standard convex-roof construction inherently preserves this condition.}: $C(\rho)$ should not increase under mixing of quantum states, i.e., $C(\sum_{\ell}p_{\ell}\rho_{\ell})\leq\sum_{\ell}p_{\ell}C(\rho_{\ell})$ for any set of states $\{\rho_{\ell}\}$ and probability $p_{\ell}\geq 0$ and $\sum_{\ell} p_{\ell}=1$.
\end{itemize}
$C(\rho)$ can be considered a valid coherence measure if and only if it satisfies the aforementioned four conditions. The $l_1$ norm of coherence and relative entropy of coherence meet these criteria and are, therefore, proper measures of coherence. Hereon, we develop these coherence monotones and express them in terms of our obtained SFO transition probabilities.

\subsection{$l_1$-norm of Quantum Coherence}

Following the resource theory, to measure the coherence in a system, we first take the \textit{$l_{1}$-norm of coherence}, which is induced by the unitarily invariant $l_{1}$ matrix norm and is given as a function of the density matrix $\rho(t)$ as \cite{Baumgratz:2014yfv,Ding:2021zfi},
\begin{equation}
     C_{l_1}\left(\rho\right)\equiv\sum_{i\neq j}|\rho_{ij}|\geq 0,
     \label{coh}
\end{equation}
where $\rho_{ij}=\langle i|\rho|j\rangle$, so this measure essentially sums up the absolute values of the off-diagonal $(i\neq j)$ elements of the density matrix $\rho$. 
We define a $d$-dimensional maximally coherent state as one that enables the deterministic generation of all other $d$-dimensional quantum states using free operations. The canonical example of a maximally coherent state is,
\begin{equation}
|\psi_d\rangle = \frac{1}{\sqrt{d}} \sum_{i=0}^{d-1} |i\rangle
\end{equation}
which has an equal amplitude for each basis set representing uniform superposition of states. 
Thus, the maximum allowed value of \(C_{l_1}\) which corresponds to the value for the maximally coherent state is $d-1$, where $d$ represents the dimension of the density matrix.
 
Now let us consider that left-handed (LH) electron neutrinos $|\nu_{e}^{L}\rangle$ of Dirac nature are produced at initial time $t=0$ and then undergo SFO. For three-flavor mixing case, the time evolution of the flavor eigenstate $|\nu_{e}^{L}\rangle$ can be expressed as \cite{Song:2018bma},
\begin{equation}
     |\nu_{e}^{L}(t)\rangle = a_{ee}^{LL}|\nu_{e}^{L}\rangle + a_{ee}^{LR}|\nu_{e}^{R}\rangle + a_{e\mu}^{LL}|\nu_{\mu}^{L}\rangle + a_{e\mu}^{LR}|\nu_{\mu}^{R}\rangle + a_{e\tau}^{LL}|\nu_{\tau}^{L}\rangle +a_{e\tau}^{LR}|\nu_{\tau}^{R}\rangle, 
     \label{cevol}
\end{equation}
where the six $a$'s are coefficients expressed in terms of the probability ($P_{e\beta}^{Lh'}$) of finding the neutrino in the states $|\nu_{e}^{L}\rangle$, $|\nu_{e}^{R}\rangle$, $|\nu_{\mu}^{L}\rangle$, $|\nu_{\mu}^{R}\rangle$, $|\nu_{\tau}^{L}\rangle$ and $|\nu_{\tau}^{R}\rangle$ respectively as,
\begin{eqnarray}
     |a_{ee}^{LL}|^2=|\langle \nu_{e}^{L}|\nu_{e}^{L} (t)\rangle|^2=P_{ee}^{LL}, \quad |a_{ee}^{LR}|^2=|\langle \nu_{e}^{R}|\nu_{e}^{L}(t)\rangle|^2=P_{ee}^{LR}, \\
     |a_{e\mu}^{LL}|^2=|\langle \nu_{\mu}^{L}|\nu_{e}^{L}(t)\rangle|^2=P_{e\mu}^{LL}, 
     \quad |a_{e\mu}^{LR}|^2=|\langle \nu_{\mu}^{R}|\nu_{e}^{L}(t)\rangle|^2=P_{e\mu}^{LR}, 
     \\
     |a_{e\tau}^{LL}|^2=|\langle \nu_{\tau}^{L}|\nu_{e}^{L}(t)\rangle|^2=P_{e\tau}^{LL}, \quad|a_{e\tau}^{LR}|^2=|\langle \nu_{\tau}^{R}|\nu_{e}^{L}(t)\rangle|^2=P_{e\tau}^{LR}.
     \label{Ptau}
\end{eqnarray}

We can hence write our $6\times 6$ density matrix $\rho_{e,3} (t)=|\nu_{e}^{L}(t)\rangle\langle\nu_{e}^{L}(t)|$ in the basis \\ $\{|\nu_{e}^{L}\rangle, |\nu_{e}^{R}\rangle, |\nu_{\mu}^{L}\rangle, |\nu_{\mu}^{R}\rangle, |\nu_{\tau}^{L}\rangle, |\nu_{\tau}^{R}\rangle \}$ as,
\begin{equation}
    \rho_{e,3} (t)= \begin{pmatrix}
    |a_{ee}^{LL}(t)|^2 & a_{ee}^{\ast LR}a_{ee}^{LL} & a_{e\mu}^{\ast LL}a_{ee}^{LL} & a_{e\mu}^{\ast LR}a_{ee}^{LL} & a_{e\tau}^{\ast LL}a_{ee}^{LL} & a_{e\tau}^{\ast LR}a_{ee}^{LL}\\
    a_{ee}^{\ast LL}a_{ee}^{LR} & |a_{ee}^{LR}(t)|^2 & a_{e\mu}^{\ast LL}a_{ee}^{LR} & a_{e\mu}^{\ast LR}a_{ee}^{LR} & a_{e\tau}^{\ast LL}a_{ee}^{LR} & a_{e\tau}^{\ast LR}a_{ee}^{LR}\\
    a_{ee}^{\ast LL}a_{e\mu}^{LL} & a_{ee}^{\ast LR}a_{e\mu}^{LL} & |a_{e\mu}^{LL}(t)|^2 & a_{e\mu}^{\ast LR}a_{e\mu}^{LL} & a_{e\tau}^{\ast LL}a_{e\mu}^{LL} & a_{e\tau}^{\ast LR}a_{e\mu}^{LL} \\
    a_{ee}^{\ast LL}a_{e\mu}^{LR} & a_{ee}^{\ast LR}a_{e\mu}^{LR} & a_{e\mu}^{\ast LL}a_{e\mu}^{LR} & |a_{e\mu}^{LR}(t)|^2 & a_{e\tau}^{\ast LL}a_{e\mu}^{LR} & a_{e\tau}^{\ast LR}a_{e\mu}^{LR}\\
    a_{ee}^{\ast LL}a_{e\tau}^{LL} & a_{ee}^{\ast LR}a_{e\tau}^{LL} & a_{e\mu}^{\ast LL}a_{e\tau}^{LL} & a_{e\mu}^{\ast LR}a_{e\tau}^{LL} & |a_{e\tau}^{LL}(t)|^2 & a_{e\tau}^{\ast LR}a_{e\tau}^{LL}\\
    a_{ee}^{\ast LL}a_{e\tau}^{LR} & a_{ee}^{\ast LR}a_{e\tau}^{LR} & a_{e\mu}^{\ast LL}a_{e\tau}^{LR} & a_{e\mu}^{\ast LR}a_{e\tau}^{LR} & a_{e\tau}^{\ast LL}a_{e\tau}^{LR} & |a_{e\tau}^{LR}(t)|^2
    \end{pmatrix}.
    \label{den}
\end{equation}

So, in our case, the maximum value of the $l_{1}$-norm of coherence is $5$ as the dimension ($d$) of our density matrix $\rho_{e,3}$ is $6$.
Using Eq. \eqref{den} and the definition in Eq. \eqref{coh}, the $l_1$-norm of coherence can thus be expressed as,
{\small\begin{eqnarray}
    C_{l_1,e}^{3f} &=& 2\big(|a_{ee}^{LR}a_{ee}^{LL}|+|a_{e\mu}^{LL}a_{ee}^{LL}|+|a_{e\mu}^{LR}a_{ee}^{LL}|+|a_{e\mu}^{LL}a_{ee}^{LR}|+|a_{e\mu}^{LR}a_{ee}^{LR}|+|a_{e\mu}^{LR}a_{e\mu}^{LL}|+|a_{e\tau}^{LL}a_{ee}^{LL}| + |a_{e\tau}^{LR}a_{ee}^{LL}|\nonumber \\ &+& |a_{e\tau}^{LL}a_{ee}^{LR}|+|a_{e\tau}^{LR}a_{ee}^{LR}|+|a_{e\tau}^{LL}a_{e\mu}^{LL}|+|a_{e\tau}^{LR}a_{e\mu}^{LL}|+|a_{e\tau}^{LL}a_{e\mu}^{LR}|+|a_{e\tau}^{LR}a_{e\mu}^{LR}|+|a_{e\tau}^{LR}a_{e\tau}^{LL}|\big), \label{coh-a}   
\end{eqnarray}}
or in terms of the transition probabilities ($P_{e\beta}^{Lh'}$) as,
\begin{eqnarray}
    C_{l_1,e}^{3f} &=& 2\bigg(\sqrt{P_{ee}^{LR}P_{ee}^{LL}}+\sqrt{P_{e\mu}^{LL}P_{ee}^{LL}}+\sqrt{P_{e\mu}^{LR}P_{ee}^{LL}}+\sqrt{P_{e\mu}^{LL}P_{ee}^{LR}}+\sqrt{P_{e\mu}^{LR}P_{ee}^{LR}} \nonumber \\ &+& \sqrt{P_{e\mu}^{LR}P_{e\mu}^{LL}} + \sqrt{P_{e\tau}^{LL}P_{ee}^{LL}} + \sqrt{P_{e\tau}^{LR}P_{ee}^{LL}} + \sqrt{P_{e\tau}^{LL}P_{ee}^{LR}} + \sqrt{P_{e\tau}^{LR}P_{ee}^{LR}} \nonumber \\ &+& \sqrt{P_{e\tau}^{LL}P_{e\mu}^{LL}} + \sqrt{P_{e\tau}^{LR}P_{e\mu}^{LL}} +\sqrt{P_{e\tau}^{LL}P_{e\mu}^{LR}} + \sqrt{P_{e\tau}^{LR}P_{e\mu}^{LR}} + \sqrt{P_{e\tau}^{LR}P_{e\tau}^{LL}}\bigg) \,.
    \label{coh-P}
\end{eqnarray}
We can now plug in the probabilities from Eqs. \eqref{pp1f} and \eqref{pp2f} into this Eq. \eqref{coh-P} to get the $l_1$-norm of coherence.

If we considered the initial state prepared at $t=0$ to be $|\nu_\mu^L\rangle$, then its time evolution would be
\begin{equation}
     |\nu_{\mu}^{L}(t)\rangle=a_{\mu e}^{LL}|\nu_{e}^{L}\rangle + a_{\mu e}^{LR}|\nu_{e}^{R}\rangle + a_{\mu\mu}^{LL}|\nu_{\mu}^{L}\rangle + a_{\mu\mu}^{LR}|\nu_{\mu}^{R}\rangle + a_{\mu\tau}^{LL}|\nu_{\tau}^{L}\rangle + a_{\mu\tau}^{LR}|\nu_{\tau}^{R}\rangle,
     \label{cevolmu}
\end{equation}
where again the six $a$'s here are coefficients which can be expressed in terms of the SFO probability ($P_{\mu\beta}^{Lh'}$) of finding the neutrino in the states $|\nu_{e}^{L}\rangle$, $|\nu_{e}^{R}\rangle$, $|\nu_{\mu}^{L}\rangle$, $|\nu_{\mu}^{R}\rangle$, $|\nu_{\tau}^{L}\rangle$ and $|\nu_{\tau}^{R}\rangle$ respectively.
Thus, density matrix $\rho_{\mu,3} (t)=|\nu_{\mu}^{L}(t)\rangle\langle\nu_{\mu}^{L}(t)|$ takes the form,
\begin{equation}
    \rho_{\mu,3} (t)= \begin{pmatrix}
    |a_{\mu e}^{LL}(t)|^2 & a_{\mu e}^{\ast LR}a_{\mu e}^{LL} & a_{\mu \mu}^{\ast LL}a_{\mu e}^{LL} & a_{\mu \mu}^{\ast LR}a_{\mu e}^{LL} & a_{\mu \tau}^{\ast LL}a_{\mu e}^{LL} & a_{\mu \tau}^{\ast LR}a_{\mu e}^{LL}\\
    a_{\mu e}^{\ast LL}a_{\mu e}^{LR} & |a_{\mu e}^{LR}(t)|^2 & a_{\mu \mu}^{\ast LL}a_{\mu e}^{LR} & a_{\mu \mu}^{\ast LR}a_{\mu e}^{LR} & a_{\mu \tau}^{\ast LL}a_{\mu e}^{LR} & a_{\mu \tau}^{\ast LR}a_{\mu e}^{LR}\\
    a_{\mu e}^{\ast LL}a_{\mu \mu}^{LL} & a_{\mu e}^{\ast LR}a_{\mu \mu}^{LL} & |a_{\mu \mu}^{LL}(t)|^2 & a_{\mu \mu}^{\ast LR}a_{\mu \mu}^{LL} & a_{\mu \tau}^{\ast LL}a_{\mu \mu}^{LL} & a_{\mu \tau}^{\ast LR}a_{\mu \mu}^{LL} \\
    a_{\mu e}^{\ast LL}a_{\mu \mu}^{LR} & a_{\mu e}^{\ast LR}a_{\mu \mu}^{LR} & a_{\mu \mu}^{\ast LL}a_{\mu \mu}^{LR} & |a_{\mu \mu}^{LR}(t)|^2 & a_{\mu \tau}^{\ast LL}a_{\mu \mu}^{LR} & a_{\mu \tau}^{\ast LR}a_{\mu \mu}^{LR}\\
    a_{\mu e}^{\ast LL}a_{\mu \tau}^{LL} & a_{\mu e}^{\ast LR}a_{\mu \tau}^{LL} & a_{\mu \mu}^{\ast LL}a_{\mu \tau}^{LL} & a_{\mu \mu}^{\ast LR}a_{\mu \tau}^{LL} & |a_{\mu \tau}^{LL}(t)|^2 & a_{\mu \tau}^{\ast LR}a_{\mu \tau}^{LL}\\
    a_{\mu e}^{\ast LL}a_{\mu \tau}^{LR} & a_{\mu e}^{\ast LR}a_{\mu \tau}^{LR} & a_{\mu \mu}^{\ast LL}a_{\mu \tau}^{LR} & a_{\mu \mu}^{\ast LR}a_{\mu \tau}^{LR} & a_{\mu \tau}^{\ast LL}a_{\mu \tau}^{LR} & |a_{\mu \tau}^{LR}(t)|^2
    \end{pmatrix}.
    \label{denm}
\end{equation}

Hence, following a similar procedure as before, the expression for the $l_1$-norm of coherence would change to,
\begin{eqnarray}
    C_{l_1,\mu}^{3f} &=& 2\bigg(\sqrt{P_{\mu e}^{LR}P_{\mu e}^{LL}}+\sqrt{P_{\mu \mu}^{LL}P_{\mu e}^{LL}}+\sqrt{P_{\mu \mu}^{LR}P_{\mu e}^{LL}}+\sqrt{P_{\mu \mu}^{LL}P_{\mu e}^{LR}}+\sqrt{P_{\mu \mu}^{LR}P_{\mu e}^{LR}} \nonumber \\ &+& \sqrt{P_{\mu \mu}^{LR}P_{\mu \mu}^{LL}} + \sqrt{P_{\mu \tau}^{LL}P_{\mu e}^{LL}} + \sqrt{P_{\mu \tau}^{LR}P_{\mu e}^{LL}} + \sqrt{P_{\mu \tau}^{LL}P_{\mu e}^{LR}} + \sqrt{P_{\mu \tau}^{LR}P_{\mu e}^{LR}} \nonumber \\ &+& \sqrt{P_{\mu \tau}^{LL}P_{\mu \mu}^{LL}} + \sqrt{P_{\mu \tau}^{LR}P_{\mu \mu}^{LL}} +\sqrt{P_{\mu \tau}^{LL}P_{\mu \mu}^{LR}} + \sqrt{P_{\mu \tau}^{LR}P_{\mu \mu}^{LR}} + \sqrt{P_{\mu \tau}^{LR}P_{\mu \tau}^{LL}}\bigg) \,.
    \label{coh-Pm}
\end{eqnarray}
We can thus plug the probabilities from Eqs. \eqref{pp1fm} and \eqref{pp2fm} into Eq. \eqref{coh-Pm} to obtain the $l_1$-norm of coherence.

In the case of two-flavor mixing, for the evolution of the initial state $|\nu_e^L\rangle$, the last two terms in Eq. \eqref{cevol} vanish, and consequently, the last two probabilities in Eq. \eqref{Ptau} also vanish. The density matrix $\rho_{e,2} (t)=|\nu_{e}^{L}(t)\rangle\langle\nu_{e}^{L}(t)|$ becomes $4\times 4$  in the basis $\{ |\nu_{e}^{L}\rangle, |\nu_{e}^{R}\rangle, |\nu_{\mu}^{L}\rangle, |\nu_{\mu}^{R}\rangle \}$ as,
\begin{equation}
    \rho_{e,2} (t)= \begin{pmatrix}
    |a_{ee}^{LL}(t)|^2 & a_{ee}^{\ast LR}a_{ee}^{LL} & a_{e\mu}^{\ast LL}a_{ee}^{LL} & a_{e\mu}^{\ast LR}a_{ee}^{LL} \\
    a_{ee}^{\ast LL}a_{ee}^{LR} & |a_{ee}^{LR}(t)|^2 & a_{e\mu}^{\ast LL}a_{ee}^{LR} & a_{e\mu}^{\ast LR}a_{ee}^{LR} \\
    a_{ee}^{\ast LL}a_{e\mu}^{LL} & a_{ee}^{\ast LR}a_{e\mu}^{LL} & |a_{e\mu}^{LL}(t)|^2 & a_{e\mu}^{\ast LR}a_{e\mu}^{LL} \\
    a_{ee}^{\ast LL}a_{e\mu}^{LR} & a_{ee}^{\ast LR}a_{e\mu}^{LR} & a_{e\mu}^{\ast LL}a_{e\mu}^{LR} & |a_{e\mu}^{LR}(t)|^2
    \end{pmatrix}.
    \label{den-2}
\end{equation}

In a similar manner to the three-flavor mixing case above, we can get the $l_1$-norm of coherence $C_{l_1,e}^{2f}$ in terms of the transition probabilities ($P_{e\beta}^{Lh'}$) for the two-flavor case as,
\begin{equation}
C_{l_1,e}^{2f} = 2\bigg(\sqrt{P_{ee}^{LR}P_{ee}^{LL}}
+\sqrt{P_{e\mu}^{LL}P_{ee}^{LL}}
+\sqrt{P_{e\mu}^{LR}P_{ee}^{LL}}
+\sqrt{P_{e\mu}^{LL}P_{ee}^{LR}}
+\sqrt{P_{e\mu}^{LR}P_{ee}^{LR}}
+\sqrt{P_{e\mu}^{LR}P_{e\mu}^{LL}}\bigg)\,,
\label{coh-P2} 
\end{equation}
and considering $|\nu_\mu^L\rangle$ as the initial state, $C_{l_1,\mu}^{2f}$ can be expressed in terms of the probabilities ($P_{\mu\beta}^{Lh'}$) as,
\begin{equation}
C_{l_1,\mu}^{2f} = 2\bigg(\sqrt{P_{\mu e}^{LR}P_{\mu e}^{LL}}
+\sqrt{P_{\mu \mu}^{LL}P_{\mu e}^{LL}}
+\sqrt{P_{\mu \mu}^{LR}P_{\mu e}^{LL}}
+\sqrt{P_{\mu \mu}^{LL}P_{\mu e}^{LR}}
+\sqrt{P_{\mu \mu}^{LR}P_{\mu e}^{LR}}
+\sqrt{P_{\mu \mu}^{LR}P_{\mu \mu}^{LL}}\bigg)\,.
\label{coh-P2m} 
\end{equation}
We can see here in the two-flavor mixing case, from the probabilities from Eqs. \eqref{pllee}, \eqref{plree}, \eqref{pllem}, \eqref{plrem} and their substitutions into the Eq. \eqref{coh-P2} when the initial state is considered to be $|\nu_e^L\rangle$ and into Eq. \eqref{coh-P2m} when the initial state is considered to be $|\nu_\mu^L\rangle$, the $l_1$-norms of coherence for both initial states will be the same (i.e. $C_{l_1,e}^{2f}=C_{l_1,\mu}^{2f}$).

\subsection{Relative Entropy of Quantum Coherence}

The Von Neumann entropy of a quantum state represented by a density matrix $\rho (t)$ can be written as\footnote{Hereon, $\log$ represents natural logarithm, with base $e$.},
\begin{equation}
    S(\rho)\equiv-\rm{Tr}(\rho\log{\rho})=-\sum_{i} \lambda_i \log{\lambda_i},
    \label{ent}
\end{equation}
where $\lambda_i$ denotes the eigenvalues of the density matrix $\rho (t)$.

The metric distance between the quantum state $\rho$ and its nearest incoherent state $\sigma$ is quantified by the quantum relative entropy as,
\begin{equation}
    S(\rho\|\sigma)=-\rm{Tr}\big(\rho(\log{\sigma})\big)-S(\rho)= \rm{Tr}\big(\rho(\log{\rho}-\log{\sigma})\big),
\label{rel}
\end{equation}
where $\sigma\in\mathcal{I}$ i.e. set of incoherent quantum states.
Now, using the above two Eqs. \eqref{ent} and \eqref{rel}, we can see that,
\begin{equation}
    S(\rho_D\|\sigma)-S(\rho\|\sigma)=S(\rho)-S(\rho_D),
\end{equation}
where $\rho_D (t)$ is the dephased state formed by the diagonal part of the density matrix $\rho (t)$.

Therefore, we can now turn our attention to the canonical coherence monotone called the relative entropy of coherence, which plays a crucial part in distillation of coherence, where it gets the operational interpretation as the optimal rate to distill maximally coherent state from the given state $\rho$ by incoherent operations $\Xi_{\rm ICPTP}(\rho)$ in the asymptotic limit of many copies of a state \cite{Winter:2016bkw}.
The distillable coherence or the \textit{relative entropy of coherence} is expressed in a closed-form expression as \cite{Baumgratz:2014yfv,Ding:2021zfi,Streltsov:2016iow},
\begin{equation}
    C_{\rm RE}(\rho) = S(\rho_{D}(t))-S(\rho(t)) = \rm{Tr}\big(\rho(t)\log{\rho(t)}-\rho_D(t)\log{\rho_D(t)}\big)\,.
    \label{relen}
\end{equation}
Here, the functional $C_{\rm RE}(\rho)$ satisfies not only the four conditions mentioned at the starting of this section, but also an additional condition of \textit{additivity} under tensor products, i.e., $C_{\rm RE}\left(\rho\otimes\sigma\right)=C_{\rm RE}\left(\rho\right)+C_{\rm RE}\left(\sigma\right)$.

Now our pure state density matrix $\rho_{e,3} (t)=|\nu_{e}^{L}(t)\rangle\langle\nu_{e}^{L}(t)|$ can be decomposed in the basis\\
$\{|\nu_{e}^{L}\rangle,  |\nu_{e}^{R}\rangle, |\nu_{\mu}^{L}\rangle,  |\nu_{\mu}^{R}\rangle, |\nu_{\tau}^{L}\rangle, |\nu_{\tau}^{R}\rangle \}\equiv|\nu_{\beta}^{h'}\rangle$ as,
\begin{equation}
    \rho_{e,3} (t)=\sum_{h,h'=L}^{R}\sum_{\alpha,\beta=e}^{\tau} a_{e\alpha}^{\ast Lh}a_{e\beta}^{Lh'}|\nu_{\alpha}^{h}(0)\rangle\langle\nu_{\beta}^{h'}(0)|\,,
    \label{den1}
\end{equation}
and its nearest optimal incoherent state (which is given by the dephased operator $\rho_D$) can be decomposed as,
\begin{equation}
    \rho_{De,3} (t)=\sum_{h'=L}^{R}\sum_{\beta=e}^{\tau} |a_{e\beta}^{Lh'}|^{2} |\nu_{\beta}^{h'}(0)\rangle\langle\nu_{\beta}^{h'}(0)|\,.
    \label{den2}
\end{equation}

We can write the relative entropy of coherence by opening up the trace in Eq. \eqref{relen} in the basis mentioned above for Eqs. \eqref{den1} and \eqref{den2} as,
\begin{equation}
    C_{\rm RE,e}^{3f}(\rho_{e,3})=\sum_{\{|\nu_\beta^{h'}\rangle\}}\langle \nu_\beta^{h'} (0) |(\rho_{e,3}(t)\log{\rho_{e,3}(t)} - \rho_{De,3}(t)\log{\rho_{De,3}(t)})|\nu_\beta^{h'} (0)\rangle \,.
    \label{relen-1}
\end{equation}
Putting in the density matrix decompositions from Eqs. \eqref{den1} and \eqref{den2} in the above Eq. \eqref{relen-1}, we get,
{\small\begin{eqnarray}
    C_{\rm RE,e}^{3f}(\rho_{e,3})&=&\sum_{h,h'=L}^{R}\sum_{\alpha,\beta=e}^{\tau}\bigg( \langle \nu_\beta^{h'} (0)||a_{e\alpha}^{\ast Lh}a_{e\beta}^{Lh'}||\nu_{\alpha}^{h}(0)\rangle\langle\nu_{\beta}^{h'}(0)| \log{(|a_{e\alpha}^{\ast Lh}a_{e\beta}^{Lh'}|)}|\nu_{\alpha}^{h}(0)\rangle\langle\nu_{\beta}^{h'}(0)|\nu_{\beta}^{h'}(0)\rangle \bigg) \nonumber\\
&-&\sum_{h'=L}^{R}\sum_{\beta=e}^{\tau}\bigg(\langle \nu_\beta^{h'} (0)||a_{e\beta}^{Lh'}|^{2}|\nu_{\beta}^{h'}(0)\rangle\langle\nu_{\beta}^{h'}(0)| \log{(|a_{e\beta}^{Lh'}|^{2})}|\nu_{\beta}^{h'}(0)\rangle\langle\nu_{\beta}^{h'}(0)|\nu_{\beta}^{h'}(0)\rangle \bigg)\,.
\end{eqnarray}}

Now using the orthonormality of our eigenbasis, i.e., $\langle\nu_{\beta}^{h'}(0)|\nu_{\alpha}^{h}(0)\rangle=\delta_{\alpha\beta}\delta_{hh'}$, we obtain the above expression in terms of the transition probabilities ($P_{e\beta}^{Lh'}$) as,
\begin{equation}
    C_{\rm RE,e}^{3f}(\rho_{e,3})=-\sum_{h'=L}^{R}\sum_{\beta=e}^{\tau} |a_{e\beta}^{Lh'}|^{2}|\log{(|a_{e\beta}^{Lh'}|^{2})}=-\sum_{h'=L}^{R}\sum_{\beta=e}^{\tau} P_{e\beta}^{Lh'}\log{(P_{e\beta}^{Lh'})}=S\left(\rho_{De,3}(t)\right)\,.
    \label{relen-P}
\end{equation}
This essentially showcases the \textit{uniqueness} property of the relative entropy of coherence for pure states, i.e., $C_{\rm RE}\left(|\psi\rangle\langle\psi|\right)=S\left(\rho_D\right)$.
The maximal value of the relative entropy of coherence for a pure state represented by a $d$-dimensional density operator is $\log{d}$.
Expanding the above Eq. \eqref{relen-P}, we arrive at the following result,
\begin{eqnarray}
    C_{\rm RE,e}^{3f}(\rho_{e,3})&=& -\bigg(P_{ee}^{LL}\log{\left(P_{ee}^{LL}\right)}+P_{ee}^{LR}\log{\left(P_{ee}^{LR}\right)}+P_{e\mu}^{LL}\log{\left(P_{e\mu}^{LL}\right)}+P_{e\mu}^{LR}\log{\left(P_{e\mu}^{LR}\right)} \nonumber\\
&+& \quad P_{e\tau}^{LL}\log{\left(P_{e\tau}^{LL}\right)}+P_{e\tau}^{LR}\log{\left(P_{e\tau}^{LR}\right)}\bigg)\,.
\label{relen-Pf}
\end{eqnarray}

We can now plug in the three-flavor SFO probabilities from Eqs. \eqref{pp1f} and \eqref{pp2f} into this Eq. \eqref{relen-Pf} to get the relative entropy of coherence.\\
If we considered the initial state prepared at $t=0$ to be $|\nu_\mu^L\rangle$, then the density matrix $\rho_{\mu,3} (t)=|\nu_{\mu}^{L}(t)\rangle\langle\nu_{\mu}^{L}(t)|$  can be decomposed in the basis $|\nu_{\beta}^{h'}\rangle$ as,
\begin{equation}
    \rho_{\mu,3} (t)=\sum_{h,h'=L}^{R}\sum_{\alpha,\beta=e}^{\tau} a_{\mu\alpha}^{\ast Lh}a_{\mu\beta}^{Lh'}|\nu_{\alpha}^{h}(0)\rangle\langle\nu_{\beta}^{h'}(0)|\,,
    \label{den1m}
\end{equation}
and the diagonal part of this density matrix $\rho_{\mu,3} (t)$ can be decomposed as,
\begin{equation}
    \rho_{D\mu,3} (t)=\sum_{h'=L}^{R}\sum_{\beta=e}^{\tau} |a_{\mu\beta}^{Lh'}|^{2} |\nu_{\beta}^{h'}(0)\rangle\langle\nu_{\beta}^{h'}(0)|\,.
    \label{den2m}
\end{equation}
Thus, the expression for relative entropy of coherence churns into,
\begin{equation}
    C_{\rm RE,\mu}^{3f}(\rho_{\mu,3})=-\sum_{h'=L}^{R}\sum_{\beta=e}^{\tau} |a_{\mu\beta}^{Lh'}|^{2}|\log{(|a_{\mu\beta}^{Lh'}|^{2})}=-\sum_{h'=L}^{R}\sum_{\beta=e}^{\tau} P_{\mu\beta}^{Lh'}\log{(P_{\mu\beta}^{Lh'})}=S\left(\rho_{D\mu,3}(t)\right)\,,
\label{relen-Pm}
\end{equation}

expanding this, we can write as,
\begin{eqnarray}
    C_{\rm RE,\mu}^{3f}(\rho_{\mu,3})&=& -\bigg(P_{\mu e}^{LL}\log{\left(P_{\mu e}^{LL}\right)}+P_{\mu e}^{LR}\log{\left(P_{\mu e}^{LR}\right)}+P_{\mu \mu}^{LL}\log{\left(P_{\mu \mu}^{LL}\right)}+P_{\mu \mu}^{LR}\log{\left(P_{\mu \mu}^{LR}\right)} \nonumber\\
&+& \quad P_{\mu \tau}^{LL}\log{\left(P_{\mu \tau}^{LL}\right)}+P_{\mu \tau}^{LR}\log{\left(P_{\mu \tau}^{LR}\right)}\bigg)\,.
\label{relen-Pmf}
\end{eqnarray}

Here, we can plug in the probabilities from Eqs. \eqref{pp1fm} and \eqref{pp2fm} into this Eq. \eqref{relen-Pmf} to get the relative entropy of coherence for neutrinos undergoing SFO with three-flavor mixing.\\
As for the two-flavor mixing scenario, $C_{\rm RE,e}^{2f}(\rho_{e,2})$ and $C_{\rm RE,\mu}^{2f}(\rho_{\mu,2})$ can be calculated using the same methodology as used to arrive at Eq. \eqref{relen-Pf} and Eq. \eqref{relen-Pmf} respectively, with the summation on flavor indices $\alpha,\beta$ running through only $e$ and $\mu$ as, 

{\begin{eqnarray}
    C_{\rm RE,e}^{2f}(\rho_{e,2})&=&  -\bigg(P_{ee}^{LL}\log{\left(P_{ee}^{LL}\right)}+P_{ee}^{LR}\log{\left(P_{ee}^{LR}\right)}+P_{e\mu}^{LL}\log{\left(P_{e\mu}^{LL}\right)}+P_{e\mu}^{LR}\log{\left(P_{e\mu}^{LR}\right)}\bigg)\,,\nonumber\\
\label{relen-P2f}\\
C_{\rm RE,\mu}^{2f}(\rho_{\mu,2})&=&  -\bigg(P_{\mu e}^{LL}\log{\left(P_{\mu e}^{LL}\right)}+P_{\mu e}^{LR}\log{\left(P_{\mu e}^{LR}\right)}+P_{\mu \mu}^{LL}\log{\left(P_{\mu \mu}^{LL}\right)}+P_{\mu \mu}^{LR}\log{\left(P_{\mu \mu}^{LR}\right)}\bigg)\,.\nonumber\\
\label{relen-P2fm}
\end{eqnarray}}

We can now plug in the two-flavor SFO probabilities from Eqs. \eqref{pllee}, \eqref{plree}, \eqref{pllem} and \eqref{plrem} into the Eq. \eqref{relen-P2f} when the initial state is considered to be $|\nu_e^L\rangle$ and into Eq. \eqref{relen-P2fm} when the initial state is considered to be $|\nu_\mu^L\rangle$ to get the relative entropy of coherence for the two-flavor mixing case.

Neutrino oscillation probabilities are directly measurable in neutrino oscillation experiments. Therefore, any coherence measure that can be explicitly expressed as a function of these probabilities is also accessible through experimental data. In \cite{Song:2018bma}, the authors quantified coherence for three-flavor oscillations in various neutrino reactor and accelerator setups, including Daya Bay, KamLAND, MINOS, and T2K, demonstrating the experimental determination of quantum properties for particles other than photons over the longest distances studied so far. Furthermore, in the context of astrophysical neutrino setups, the observed flavor ratio at the detector is nearly equiproportional, due to the averaging-out of the oscillations over vast astrophysical distances before reaching the detector rendering it relatively insensitive to such effects. Currently, it is not feasible to directly test coherence measures in existing astrophysical neutrino setups, such as IceCube. As a result, this restricts their capability to constrain oscillation parameters, including the neutrino magnetic moment.
We will now present an analysis of the behaviour of the coherence measures for high energy astrophysical neutrinos.

\subsection{Testing measures of Coherence for various neutrino sources}

Apart from the Sun, neutrino sources within the Milky Way are diverse and include several astrophysical and potential exotic origins. These include supernovae, cosmic ray interactions, neutron stars, black holes, and potentially dark matter. Supernovae, during the explosive death, release significant bursts of neutrinos.  High-energy cosmic rays interacting with the interstellar medium also produce neutrinos. Neutron stars and black holes, via accretion processes, emit neutrinos as well. Additionally, dark matter annihilation or decay could also serve as a possible source, contributing to the overall neutrino flux detected on Earth. These various sources span a wide range of energies. Here, we consider neutrino sources located near the center of the Milky Way \cite{Anchordoqui:2013dnh,Bai:2014kba,Anchordoqui:2016dcp,Kheirandish:2020upj,Ackermann:2022rqc,Bustamante:2023iyn} so that they can travel considerable distances ($\sim$ 10 kpcs) to observe the effects of SFO in the interstellar magnetic field, which is approximately $ \sim \mu\text{G}$ \cite{Beck:2000dc}\footnote{The interstellar magnetic field is not uniform and coherent over the entire 10 kpc distance.  However, for simplicity, we assume an average value of micro-Gauss over the entire 10 kpc.}.
We consider energies between 10 TeV and 1 PeV for muon neutrinos sourced within the Milky Way \cite{Ahlers:2015moa,IceCube:2023ame,IceCube:2023hou}. We focus on muon neutrinos because they typically exhibit high fluxes and can attain high energies.

\begin{figure*} [htb]
\centering
\includegraphics[scale=0.525]{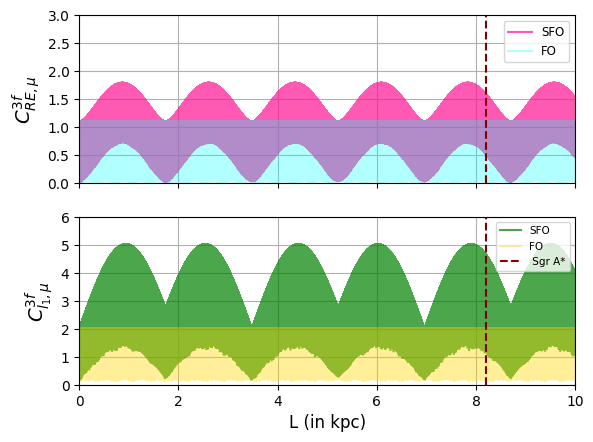} \includegraphics[scale=0.525]{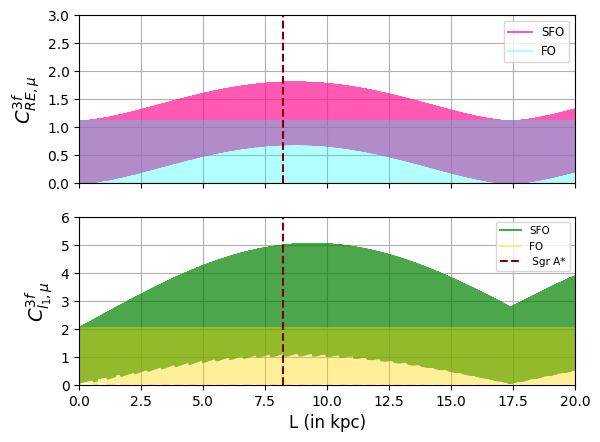}
\caption{$l_1$-norm of coherence (in green for SFO and in yellow for FO) and relative entropy of coherence (in pink for SFO and in blue for FO) with varying distance in the three-flavor mixing case for neutrino energy of 10 TeV (left) and 1 PeV (right), for the experimental upper limit $\mu_\nu\sim 10^{-12}\mu_B$ (left) and $\mu_\nu\sim 10^{-13}\mu_B$ (right) in the presence of interstellar magnetic field of the order $\sim\mu\rm G$. In both the panels, the maroon dotted vertical line denotes the location of the black hole Sagittarius $\rm A^{\ast}$ at $L_{\rm Sgr}\simeq 8.2\,\rm kpc$.}
\label{NP}
\end{figure*}

\begin{figure*} [htb]
\centering
\includegraphics[scale=0.525]{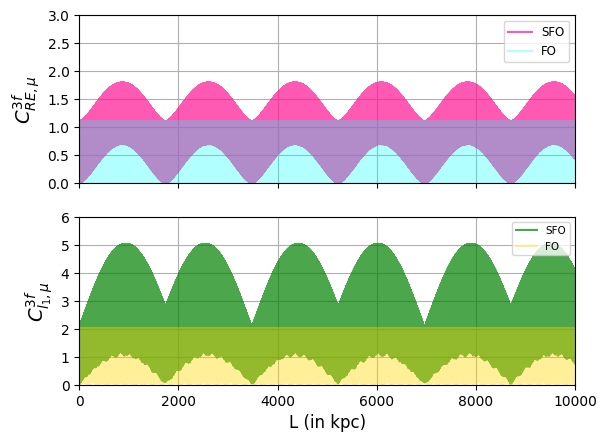} \includegraphics[scale=0.525]{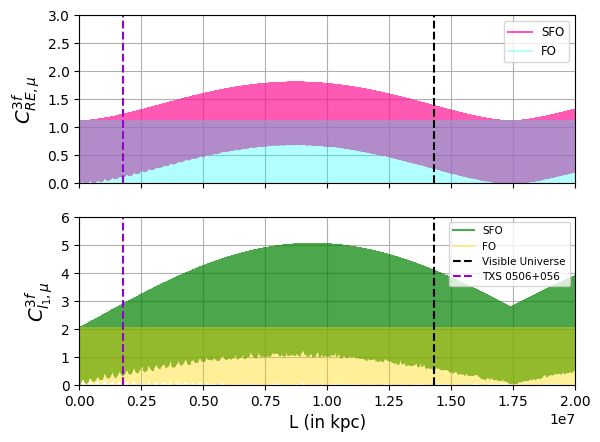}
\caption{$l_1$-norm of coherence (in green for SFO and in yellow for FO) and relative entropy of coherence (in pink for SFO and in blue for FO) with varying distance in the three-flavor mixing case for neutrino energy of $1\rm PeV$ and for the experimental upper limit $\mu_\nu\sim 10^{-12}\mu_B$ (left) and for $\mu_\nu\sim 10^{-16}\mu_B$ (right) in the presence of intergalactic magnetic field of the order $\sim\rm nG$. In the right panel, the purple dotted vertical line denotes the location of the TXS 0506+056 at $L_{\rm TXS} \simeq 1.75\,\rm Gpc$, and the black dotted vertical lines denote the size of the visible universe ($\sim 14.3\,\rm Gpc$).}
\label{NP2}
\end{figure*}

In Fig. \ref{NP} and Fig. \ref{NP2}, we have utilized the six SFO probability expressions for the three-flavor mixing scenario from Eqs. \eqref{pp1fm} and \eqref{pp2fm}, substituting them into Eq. \eqref{coh-Pm} to obtain the $l_1$-norm of coherence $C_{l_1,\mu}^{3f}$ and into Eq. \eqref{relen-Pmf} to determine the relative entropy of coherence $C_{\rm RE,\mu}^{3f}$. These measures are plotted against the propagation distance. Additionally, we compare these measures of coherence for SFO with those for standard FO in each plot.

In Fig. \ref{NP}, we consider a pervasive interstellar magnetic field of approximately $\sim \mu \rm G$ for the experimental upper limit on the neutrino magnetic moment $\mu_\nu \sim 10^{-12} \mu_B$, corresponding to a neutrino energy of 10 TeV. Additionally, we explore an order of magnitude improved upper limit of the neutrino magnetic moment $\mu_\nu \sim 10^{-13} \mu_B$, for which the neutrino energy is considered to be 1 PeV. We have also indicated the distance from the Sagittarius $\rm A^{\ast}$ black hole at the center of the Milky Way ($L_{\rm Sgr}=8178\pm13_{\rm stat}\pm22_{\rm sys}\rm pc$ \cite{Gravity:2019nxk}) with a maroon dotted vertical line on both plots.

We observe that the peaks of oscillation of the $l_1$-norm of coherence $C_{l_1,\mu}^{3f}$ fluctuate and envelop the waves for the case of SFO (plotted in green), reaching a maximum value of around $5$ at certain intervals. This value is also the theoretical maximum feasible for the $6\times 6$ density matrix, as discussed in the first subsection of this section.
In contrast, for the case of FO, the $\cos^2{(\mu_\nu B_\perp t)}$ and $\sin^2{(\mu_\nu B_\perp t)}$ terms in Eqs. \eqref{pp1fm} and \eqref{pp2fm} are absent. Consequently, the enveloping wave packing feature observed in SFO is missing here. As a result, the peaks of oscillation of $C_{l_1,\mu}^{3f}$ reach a constant maximum value of $2$, which is plotted in yellow. Obviously, this maximum value of $2$ for $C_{l_1,\mu}^{3f}$ in the case of FO is also theoretically justified since the density matrix becomes $3\times 3$ when only flavor-change is considered. This occurs because the coefficients $a_{\alpha\beta}^{LR}$ vanish in the state evolution Eqs. \eqref{cevol} and \eqref{cevolmu}, as there is no spin-flipping.
 
Owing to the cyclic behavior of probabilities, both measures of quantum coherence also exhibit cyclic variations within their allowed values. As shown in Fig. \ref{NP}, the $C_{l_1,\mu}^{3f}$ for FO completes a vast number of cycles over astrophysical distances of $\sim$ kpcs. A single cycle for FO corresponds to $\sim$ kms. This implies that FO can maintain larger values of $C_{l_1,\mu}^{3f}$ (say $\gtrsim 50\%$ of the maximum allowed value) over distances of the order of kms which are relevant for terrestrial neutrino experiments such as reactor and accelerator experiments.
In contrast, for SFO, the cycle extends to kpc-scale distances, indicating that higher values of $C_{l_1,\mu}^{3f}$ are sustained over these larger distances, relevant for astrophysical neutrinos. For instance, from the left panel, a cycle completes around 1.75 kpc, with substantial coherence maintained up to 1 kpc. Furthermore, for a 1 PeV neutrino energy and $\mu_{\nu} \sim 10^{-13}\, \mu_B$, the right panel shows that $C_{l_1,\mu}^{3f}$ does not complete a cycle even if neutrinos originate from the Milky Way's center and propagate through a distance of $L_{\rm Sgr}$. The $C_{l_1,\mu}^{3f}$ reaches a maximum value around this length scale of $L_{\rm Sgr}$. The same conclusions apply to the relative entropy of coherence.

For the relative entropy of coherence $C_{\rm RE,\mu}^{3f}$, we observe in Fig. \ref{NP} that for standard FO (plotted in blue), the peaks of oscillation consistently reach a value of around $1.1$. In the case of SFO (plotted in pink), it exhibits the expected nature of enveloping SFO waves, with the extremes of oscillation fluctuating. It reaches the maximal possible value of $\log{6}=1.792$ at certain intervals. 
From this, we can interpret that in the presence of an external magnetic field, where SFO occurs, the rate of distillation of maximally coherent neutrino states from $\rho_{\mu,3}$, quantified by $C_{\rm RE,\mu}^{3f}$, periodically increases to a much larger extent compared to the case of FO, where the periodic increase saturates at a lower constant maximal value around $\log{3}=1.098$. This is an expected maximal value theoretically, as the density operator in the case of neutrinos undergoing FO, reduces to $3\times 3$ unlike the $6$-dimensional density operator in case of SFO, for which the corresponding maximal value is $\log{6}$. Hence, the rate of distillation of coherence is expected to be periodically much more enhanced in the case of SFO, where evidently, we get the additional degrees of freedom due to the combined effect of the neutrino magnetic moment and the pervasive magnetic field allowing this enhancement in the values of the $C_{\rm RE,\mu}^{3f}$ oscillations. We can see from both plots that, unlike the case of FO, the $l_1$-norm and the relative entropy of coherence retain a finite non-zero value for a considerably large distance for the case of SFO.
These periods of fluctuation of the extremes in both measures of coherence for the case of SFO depend primarily on the combined value of the neutrino magnetic moment $\mu_\nu$ and the strength of the external magnetic field $B$.

We show that the results are nearly independent of neutrino energy, since the probability-dependent coherence measures vary with the ratio of propagation length to neutrino energy. Given the vast length scales involved, ranging from kpc to Gpc, the energy dependence of the coherence measures becomes negligible within the considered energy range. This approach of quantifying coherence on astrophysical scales for neutrinos having a magnetic moment, traveling through interstellar or intergalactic magnetic fields, can similarly be applied to neutrinos coming from any distant source and having energies of the order of MeV.

A galactic supernova at a distance of around 10 kpc from Earth is expected to produce approximately \(10^4\) \(\bar{\nu}_e\) events at Earth based detectors. As these MeV-neutrinos propagate through the dense and turbulent environment of the supernova, matter effects play a crucial role. Beyond the standard resonant flavor conversion in matter, the innermost regions of the supernova, characterized by an extremely high neutrino density, exhibit flavor evolution dominated by collective behavior induced by \(\nu-\nu\) self interactions \cite{Pantaleone:1992eq,Volpe:2023met,Dighe:2009nr,Mirizzi:2015eza}. The study of neutrinos originating from such sources necessitates a more detailed analysis of spin-flavor oscillations and hence the coherence measures in these specific environments.

In Fig. \ref{NP2}, we consider neutrino sources beyond the Milky Way, such as the TXS 0506+056 blazar, a point source producing ultrahigh-energy neutrino flux \cite{Padovani:2018acg,Kurahashi:2022utm} and located at approximately $L_{\rm TXS} \simeq 1.75$ Gpc (giga-parsecs) \cite{Keivani:2018rnh}. For these sources, intergalactic magnetic fields are considered with upper bound of $10^{-9}$ G. The predicted values of the measures of quantum coherence for these extra-galactic neutrinos are shown for 1 PeV neutrino flux with a magnetic moment $\mu_\nu \sim 10^{-12} \mu_B$ and also with a more stringent value of $\mu_\nu \sim 10^{-16} \mu_B$.
We have indicated the distance from the TXS 0506+056 blazar ($L_{\rm TXS}$) with a purple dotted vertical line on the plot in the right panel of Fig. \ref{NP2}.

As discussed earlier, the length scale of the coherence monotones for SFO, depends on the combined values of $\mu_{\nu}$, magnetic field $B$ and neutrino energy $E$. It is evident from the left panel of Fig. \ref{NP2}, that for the current upper limit on $\mu_{\nu}$, the scale of coherence measures is $\sim 1750\,\rm kpc$. On the other hand, if we consider $\mu_{\nu}\sim 10^{-16}\mu_{B}$ as in the right panel of Fig. \ref{NP2}, then the scale of coherence measures increases to $1.75\times 10^{7}\,\rm kpc$ which is way beyond the size of the visible universe (which is shown by a black dotted vertical line on the same plot). We see in the right panel of Fig. \ref{NP2} that at the length scale of the visible universe the $C_{l_1,\mu}^{3f}$ is approximately 80\% of the maximum value, and for neutrinos sourced from the TXS 0506+056 blazar propagating through a distance of $L_{\rm TXS}$, the $C_{l_1,\mu}^{3f}$ reaches around 50\% of the maximum value. For a maximally coherent state, the coherence measures reach their highest possible value because the neutrino transition and survival probabilities are balanced in such a way, allowing maximum superposition of the basis states. This balance ensures that the state exhibits the greatest degree of coherence maintained up to a certain distance periodically. In the case of SFO, for the considered \(\mu_\nu B\), Fig. \ref{NP} and Fig. \ref{NP2} demonstrate that a maximally coherent state can be attained for both coherence measures for neutrinos traveling astrophysical distances of orders exceeding kiloparsecs. This outcome is not necessarily true for FO, particularly at distance to energy ratios relevant for terrestrial ranges.\\ 

\section{Conclusions}
\label{sec:con}

Quantum coherence, representing the superposition of orthogonal states, is a fundamental concept in quantum mechanics and is precisely defined in quantum resource theory. Additionally, coherence is intrinsically linked to various measures of quantum correlations, making it a crucial concept for quantifying the inherent quantumness of a given system. Thus, analyzing quantum coherence is essential not only for understanding the foundational aspects of quantum mechanics but also for exploring the system's potential to perform various quantum information tasks.

In the context of neutrino FO, the study of quantum correlations is anticipated to be crucial, as neutrino beams undergoing FO represent superpositions of states rather than single states. This superposition is essential for the interference effects that underlie the oscillation phenomenon, making the coherence of these states fundamental to accurately describing and understanding neutrino behavior. This understanding will be crucial for exploring the potential of these systems for various applications related to quantum technologies. Such studies were previously conducted where coherence in neutrino FO was quantified using the $l_1$ norm. However, the experimental confirmation of neutrino oscillations implies that neutrinos must possess mass, which may result in the possibility of neutrinos having a magnetic dipole moment. This magnetic moment can induce SFO when neutrinos encounter external magnetic fields, leading to transitions from active to sterile neutrinos. Therefore, it will be intriguing to observe how the pattern of quantum coherence is modified when neutrinos, having additional degrees of freedom, oscillate under the influence of an external magnetic field. 

In this work, we investigate the quantum coherence in neutrino SFO, quantified by the $l_1$ norm and the relative entropy of coherence, for $\nu_\mu$ Dirac neutrino sources within and beyond the Milky Way. Our findings are listed below:
\begin{itemize}
\item The $l_1$ norm and the relative entropy of coherence can be expressed in terms of neutrino SFO probabilities. 
\item For FO, coherence measures can maintain high values (say within 50\% of the maximum) over several kilometers, relevant for terrestrial experiments like reactors and accelerators.
\item For SFO, the coherence scale extends to astrophysical distances, ranging from kiloparsecs to gigaparsecs. This scale depends primarily on the values of the neutrino magnetic moment, the strength of the interstellar or intergalactic magnetic field, and the neutrino energy.

    \item For neutrinos sourced near the center of the Milky Way and having energies between 10 TeV and 1 PeV, the scale of coherence is of the order of kpcs for values of the neutrino magnetic moment close to the current experimental upper bound of $10^{-12}$ $\mu_B$. Specifically, for 10 TeV energy and $\mu_\nu \sim 10^{-12} \mu_B$, a cycle completes around 1.75 kpc, with substantial coherence maintained up to 1 kpc. In contrast, for 1 PeV energy and $\mu_\nu \sim 10^{-13} \mu_B$, the measures of quantum coherence attain their peak value at the length scale corresponding to the location of the black hole Sagittarius $\rm A^{\ast}$  from Earth, which is approximately 8.2 kpc.

    \item For neutrinos sourced beyond the Milky Way, the scale of coherence can extend up to a few Gpcs for a neutrino energy of 1 PeV, $\mu_\nu \sim 10^{-12} \mu_B$ and an intergalactic magnetic field of the order of nG, which is the current upper bound on the intergalactic magnetic field. Keeping all other parameters the same and using a value of $\mu_\nu$ that is four orders of magnitude above the current bound, the cycle of coherence may even extend beyond the size of the visible universe. For these parameters, neutrinos propagating from the TXS 0506+056 blazar can only attain around 50\% of the maximum allowed values for the $l_1$ norm and relative entropy of coherence.

\end{itemize} 

Thus, we see that the additional degrees of freedom embedded in the neutrino system, owing to its mass and its intricate interaction with external magnetic fields due to a finite magnetic moment, enrich the embedded quantum coherence in the system. This coherence can now extend to astrophysical scales, enabling any future quantum technologies at that scale, such as communications using neutrinos within and beyond the Milky Way.

\section{Acknowledgements}

This work is a homage to our dear friend and colleague Ashutosh K. Alok whose untimely demise has left an irreplacable void in our hearts. His invaluable insights, enthusiasm and his memories will always continue to inspire us forever.
NRSC would like to express sincere gratitude to the Department of Physics ``E.R. Caianiello" and INFN, Salerno for their generous hospitality and support during the completion of the work. SG acknowledges support from the Department of Science and Technology
(DST), Government of India, under Grant No. IFA22-PH 296 (INSPIRE
Faculty Award).
GL thanks COST Action COSMIC WISPers CA21106, supported by COST (European Cooperation in Science and Technology).

\end{document}